\documentclass[prd,aps,eqsecnum]{revtex4-1}

\usepackage{hyperref,amsfonts,amsmath,amssymb,graphicx,bm,subfigure}

\begin{document}

\author{Maxim Dvornikov}
\email{maxdvo@izmiran.ru}

\title{Neutrino flavor oscillations in stochastic gravitational waves}

\affiliation{Pushkov Institute of Terrestrial Magnetism, Ionosphere
and Radiowave Propagation (IZMIRAN),
108840 Moscow, Troitsk, Russia}

%
%
%
%
%
%
%

\begin{abstract}
We study neutrino flavor oscillations in a plane gravitational wave
(GW) with circular polarization. For this purpose we use the solution
of the Hamilton-Jacobi equation to get the contribution of GW to the
effective Hamiltonian for the neutrino mass eigenstates. Then, considering
stochastic GWs, we derive the equation for the density matrix
for flavor neutrinos and analytically solve it in the two flavors
approximation. The equation for the density matrix for the three neutrino flavors is also derived and solved numerically. In both cases of two and three neutrino flavors, we predict the ratios of fluxes of different flavors
at a detector for cosmic neutrinos with relatively low energies owing
to the interaction with such a GW background. The obtained results
are compared with the recent observation of the flavor content of the astrophysical
neutrino fluxes.
\end{abstract}

\maketitle

\section{Introduction}

After the recent success in experimental studies of solar~\cite{Ago18},
atmospheric~\cite{Abe18}, and accelerator neutrinos~\cite{Ace18},
these particles are believed to be massive and have nonzero mixing
between different flavors. These neutrino properties result in transformations
of the flavor content of a neutrino beam, called neutrino flavor oscillations~\cite{Bil18}.

Neutrino flavor oscillations can happen in vacuum. However, external
fields, such as a magnetic field~\cite{Giu16} or the electroweak interaction
with background matter~\cite{MalSmi16}, can significantly modify
the process of neutrino oscillations. The neutrino interaction with
a gravitational field, in spite of its weakness, was reported in Refs.~\cite{AhlBur96,For97}
to contribute the propagation and oscillations of flavor neutrinos.
Recently, the method to account for the contribution of gravity, developed
in Refs.~\cite{AhlBur96,For97}, was further used in Refs.~\cite{GodPas11,Vis15}
to study neutrino flavor oscillations in static gravitational backgrounds
such as Schwarzchild and Kerr metrics.

It is interesting to analyze how neutrino flavor oscillations proceed
in a time dependent gravitational field, such as a gravitational wave
(GW). This interest is inspired by the recent detection of GW emitted
by merging binary black holes (BHs), reported in Ref.~\cite{Abb16}.
Later on, multiple GW signals, including that of the neutron stars coalescence,
were observed. The summary of these events is presented in Ref.~\cite{Abb18}.
These phenomena are a strong evidence of the validity of the general
relativity. Moreover, presently significant efforts are made to detect
GW and astrophysical neutrinos, emitted by the same source (see, e.g., Ref.~\cite{Alb19}). It would open a window for the multimessenger
astronomy.

The influence of GW on neutrino oscillations was studied in Ref.~\cite{Dvo19},
where spin oscillations are considered. In that situation, transitions
between left and right polarized particles, belonging to the same
neutrino type, were discussed. Neutrino spin oscillations driven by
GW, studied in Ref.~\cite{Dvo19}, are analyzed using the formalism
for the description of the neutrino spin evolution in external fields
in curved space-time developed in Refs.~\cite{Dvo06,Dvo13}. Note
that the evolution of a spinning particle in GW was also considered
in Ref.~\cite{ObuSilTer17}.

This paper, where we continue our study of neutrino oscillations in
GW, is organized in the following way. First, in Sec.~\ref{sec:FLOSCGW},
we adapt the formalism for the description of neutrino flavor oscillations,
developed in Refs.~\cite{AhlBur96,For97,GodPas11,Vis15}, to describe
the neutrino interaction with a time dependent gravitational field,
such as GW. Using the solution of the Hamilton-Jacobi equation for
a massive particle in GW, previously obtained in Ref.~\cite{Pop06},
we derive the effective Hamiltonian for neutrino flavor oscillations
in GW. Then, considering a stochastic GW background, we obtain the
equation for the neutrino density matrix and analytically solve it
for the two neutrinos system. The equation for the density matrix of three flavor neutrinos is derived and solved numerically.
In Sec.~\ref{sec:ASTROPHYS}, we consider
a possible astrophysical application consisting in the interaction
of cosmic neutrinos with random GWs emitted by coalescing supermassive BHs both in the two flavors approximation and in the general situation of the three neutrino flavors. We
predict a specific flavor content of astrophysical neutrinos in
a detector owing to the interaction with such GWs. In Sec.~\ref{sec:COMP}, we consider other random factors which can influence flavor oscillations of cosmic neutrinos on large distances.
Our results are
summarized in Sec.~\ref{sec:CONCL}. In Appendix~\ref{sec:AVERAGE},
we calculate the averaged phase of neutrino oscillations induced by
stochastic GWs. 

\section{Neutrino flavor oscillations in a gravitational wave\label{sec:FLOSCGW}}

In this section, we study flavor oscillations of neutrinos in a plain
GW. Then we assume that there is a stochastic background of GWs. The
equation for the density matrix of mixed neutrinos, accounting for
both the vacuum term and the interaction with GW, is derived and solved
analytically. Then we also consider the general case of three flavor neutrinos and present the evolution equation for the density matrix in this situation.

We start with the discussion of the system of flavor neutrinos in the two flavors approximation,
$(\nu_{e},\nu_{x})$, with $x=\mu,\tau$. These neutrino flavor eigenstates
participate in the electroweak interaction with other leptons. However,
these particles do not have definite masses. To diagonalize the mass
matrix we introduce the neutrino mass eigenstates $\psi_{a}$, $a=1,2$,
which are related to $\nu_{\lambda}$ by means of the matrix transformation
$\nu_{\lambda}=U_{\lambda a}\psi_{a}$, where $U_{\lambda a}$ are
the components of the mixing matrix. If we consider two flavor neutrinos,
$(U_{\lambda a})$ has the form,
\begin{equation}\label{eq:U}
  (U_{\lambda a})=
  \left(
    \begin{array}{cc}
      \cos\theta & \sin\theta\\
      -\sin\theta & \cos\theta
    \end{array}
  \right),
\end{equation}
where $\theta$ is the vacuum mixing angle.

The neutrino mass eigenstate with the mass $m_{a}$ was found in Refs.~\cite{AhlBur96,For97}
to evolve in a gravitational field as
\begin{equation}\label{eq:psiaSa}
  \psi_{a}(\mathbf{x},t)\sim\exp
  \left[
    -\mathrm{i}S_{a}(\mathbf{x},t)
  \right],
\end{equation}
where $S_{a}(\mathbf{x},t)$ is the action for this particle, which
obeys the Hamilton-Jacobi equation~\cite{GodPas11,Vis15},
\begin{equation}\label{eq:HJeq}
  g_{\mu\nu}\frac{\partial S_{a}}{\partial x_{\mu}}\frac{\partial S_{a}}{\partial x_{\nu}} =
  m_{a}^{2},
\end{equation}
where $g_{\mu\nu}$ is the metric tensor.

Instead of dealing with the neutrino wave functions in Eq.~(\ref{eq:psiaSa}),
we can define the contribution to the effective Hamiltonian $H_{m}$
for the mass eigenstates as
\begin{equation}\label{eq:massH}
  \left(
    H_{m}
  \right)_{aa}=\frac{\partial}{\partial t}S_{a}(|\mathbf{x}|\approx t,t),
\end{equation}
where we take that neutrinos are ultrarelativistic particles. Equation~(\ref{eq:massH})
means that $\psi_{a}$ obeys the equation, $\mathrm{i}\dot{\psi}_{a}=(H_{m})_{aa}\psi_{a}$.
One can check that, in case of two neutrino eigenstates in vacuum,
Eq.~(\ref{eq:massH}) results in the correct vacuum oscillations
phase $\Phi_{\mathrm{vac}}=\Delta m^{2}/4E$, where $\Delta m^{2}=m_{2}^{2}-m_{1}^{2}>0$
and $E$ is the mean neutrino energy.

As a background gravitational field, we consider a plane GW with the
circular polarization propagating along the $z$ axis. Choosing the
transverse-traceless gauge, we get that the metric has the form~\cite{Buo07},
\begin{equation}\label{eq:metric}
  \mathrm{d}s^{2}=
  g_{\mu\nu}\mathrm{d}x^{\mu}\mathrm{d}x^{\nu}=
  \mathrm{d}t^{2}-
  \left(
    1-h\cos\phi
  \right)
  \mathrm{d}x^{2}-
  \left(
    1+h\cos\phi
  \right)
  \mathrm{d}y^{2}+2h\sin\phi \, \mathrm{d}x\mathrm{d}y-\mathrm{d}z^{2},
\end{equation}
where $h$ is the dimensionless amplitude of the wave, $\phi=\left(\omega t-kz\right)$
is the phase of the wave, $\omega$ is frequency of the wave, and
$k$ is the wave vector. In Eq.~(\ref{eq:metric}), we use the Cartesian
coordinates $x^{\mu}=(t,x,y,z)$.

The solution of Eq.~(\ref{eq:HJeq}) for a plane GW of the arbitrary
form, not necessarily a monochromatic one as in Eq.~(\ref{eq:metric}),
was found in Ref.~\cite{Pop06}. If we define $g_{\mu\nu}^{\perp}=g_{\mu\nu}$,
at $\mu,\nu=1,2$, and
\begin{equation}
  G_{\mu\nu}=\int_{0}^{u}g_{\mu\nu}^{\perp}(u)\mathrm{d}u,
  \quad
  u=t-z,
\end{equation}
this solution takes the form,
\begin{equation}\label{eq:Sa}
  S_{a}=\frac{1}{2\mathcal{E}}
  \left[
    m_{a}^{2}u-G_{\mu\nu}p_{\perp}^{\mu}p_{\perp}^{\nu}
  \right] +
  \frac{1}{2}\mathcal{E}v+p_{\perp}^{\mu}x_{\mu}^{\perp},
\end{equation}
where $\mathcal{E}=p_{0}+p_{3}$ and $p_{\perp}^{\mu}=(0,p^{1},p^{2},0)$
are the integrals of motion of Eq.~(\ref{eq:HJeq}), $x_{\perp}^{\mu}=(0,x^{1},x^{2},0)$,
and $v=t+z$.

One can see in Eq.~(\ref{eq:Sa}), in case of a neutrino propagating
along GW, i.e. when $p_{\perp}^{\mu}=0$, this background gravitational
field does not affect the neutrino motion. If a neutrino interacts
with a plane electromagnetic wave, there is an effect on neutrino
oscillations in case when particles move along the wave~\cite{Dvo18}.
The fact that GW cannot induce a spin flip of a neutrino propagating
along the wave, i.e. it does not directly influence neutrino spin
oscillations, was revealed in Ref.~\cite{Dvo19}.

If we study the neutrino motion along GW and, then, adiabatically
turn off the gravitational field, the action in Eq.~(\ref{eq:Sa})
takes the form, $S_{a}=p_{0}x^{0}+p_{3}x^{3}$, where $p_{0}=\sqrt{m_{a}^{2}+p_{3}^{2}}$.
It means that Eq.~(\ref{eq:Sa}) has a correct vacuum limit.

Since typically $h\ll1$ in Eq.~(\ref{eq:metric}), we can decompose
$S_{a}$ in Eq.~(\ref{eq:Sa}) in a series,
\begin{equation}\label{eq:Sexpansion}
  S_{a}(h,\dots)=S_{a}^{(0)}(\dots)+hS_{a}^{(1)}(\dots)+\mathcal{O}(h^{2}),
\end{equation}
and keep only the terms linear in $h$. The symbol ``$\dotsc$'' in
the arguments of $S_{a}^{(i)}(\dots)$ incorporates other parameters
except the contribution of the gravitational interaction. Since the
amplitude of GW has been explicitly written in Eq.~(\ref{eq:Sexpansion}),
i.e. the effects of a nontrivial geometry have been taken into account,
we can conclude that the additional parameters, marked by the ``$\dotsb$'' symbol,
obey relations in a flat space-time. For example, $S_{a}^{(0)}(\dots)$
in Eq.~(\ref{eq:Sexpansion}) is the action of a free massive particle
in Minkowski space.

We suppose that a neutrino propagates arbitrarily with respect to
GW. Hence the components of the neutrino momentum have the form, $p^{1}=p\cos\varphi\sin\vartheta$,
$p^{2}=p\sin\varphi\sin\vartheta$, and $p^{3}=p\cos\vartheta$, where $\varphi$ and $\vartheta$ are the spherical angles. Using
Eqs.~(\ref{eq:massH})-(\ref{eq:Sa}), we get the diagonal entries
of $H_{m}$, which contain the linear contribution of GW, in the form,
\begin{equation}\label{eq:Haag}
  (H_{m}^{(g)})_{aa}=-\frac{p^{2}h}{2E_{a}}\sin^{2}\vartheta\cos(2\varphi-\phi_{a}),
\end{equation}
where $E_{a}=\sqrt{m_{a}^{2}+p^{2}}$ is the neutrino energy, $\phi_{a}=\omega t(1-\beta_{a}\cos\vartheta)$
is the phase of GW accounting for the fact that a neutrino moves on
a certain trajectory, which is a straight line approximately, and
$\beta_{a}=p/E_{a}$ is the neutrino velocity. Note that $(H_{m}^{(g)})_{aa}$
in Eq.~(\ref{eq:Haag}) corresponds to $hS_{a}^{(1)}(\dots)$ in
Eq.~(\ref{eq:Sexpansion}).

Now we turn to the description of the evolution of the neutrino flavor
eigenstates accounting for the contribution of GW. Using Eq.~(\ref{eq:U}),
one has that neutrino flavor eigenstates obey the Schr\"odinger equation
$\mathrm{i}\dot{\nu}_{\lambda}=(H_{f})_{\lambda\kappa}\nu_{\kappa}$,
where the effective Hamiltonian takes the form, $H_{f}=UH_{m}U^{\dagger}$.

We have taken into account the contribution of GW linear in $h$ to
the diagonal elements of $H_{m}$. However, besides GW, there are
usual vacuum contributions to these elements, which have the form
$(H_{m}^{(\mathrm{vac})})_{aa}=m_{a}^{2}/2E$. Accounting for $(H_{m}^{(\mathrm{vac})})_{aa}$,
as well as using Eqs.~(\ref{eq:U}) and~(\ref{eq:Haag}), we obtain
that $H_{f}$ has the form,
\begin{align}\label{eq:Hf}
  H_{f} & =H_{0}+H_{1},\quad H_{0}=\Phi_{\mathrm{vac}}M,
  \quad
  H_{1}=-\delta\Phi_{g}M,
  \nonumber
  \\
  M & =(\bm{\sigma}\mathbf{n}),\quad\mathbf{n}=(\sin2\theta,0,-\cos2\theta),
\end{align}
where $\delta\Phi_{g}=\left[(H_{m}^{(g)})_{11}-(H_{m}^{(g)})_{22}\right]/2$
is the contribution of GW to neutrino flavor oscillations, $(H_{m}^{(g)})_{aa}$
is given in Eq.~(\ref{eq:Haag}), and $\bm{\sigma}$ are the Pauli
matrices.

We mentioned above that a neutrino, propagating along GW, is not affected
by such GW. Therefore we consider the interaction of a neutrino with
a stochastic GW background. In this situation, following Ref.~\cite{LorBal94},
it is more convenient to deal with the density matrix $\rho$. Let
us define $\rho_{\mathrm{I}}=U_{0}^{\dagger}\rho U_{0}$, where
\begin{equation}\label{eq:U0}
  U_{0}=\exp
  \left(
    -\mathrm{i}H_{0}t
  \right) =
  \cos
  \left(
    \Phi_{\mathrm{vac}}t
  \right)-
  \mathrm{i}(\bm{\sigma}\mathbf{n})\sin
  \left(
    \Phi_{\mathrm{vac}}t
  \right).
\end{equation}
We should average $\rho_{\mathrm{I}}$ over the directions of the
GW propagation and its amplitude. Then we consider the $\delta$-correlated
Gaussian distribution of $h$: $\left\langle h(t_{1})h(t_{2})\right\rangle =2\tau\delta(t_{1}-t_{2})\left\langle h^{2}\right\rangle $,
where $\tau$ is the correlation time. The evolution equation for
$\left\langle \rho_{\mathrm{I}}\right\rangle $, obtained in Ref.~\cite{LorBal94},
has the form,
\begin{equation}\label{eq:rhoIeq}
  \frac{\mathrm{d}}{\mathrm{d}t}
  \left\langle
    \rho_{\mathrm{I}}
  \right\rangle = -
  \left\langle
    \delta\Phi_{g}^{2}
  \right\rangle
  \tau
  \left[
    \left\langle
      \rho_{\mathrm{I}}
    \right\rangle -
    M
    \left\langle
      \rho_{\mathrm{I}}
    \right\rangle
    M
  \right],
\end{equation}
where
\begin{equation}\label{eq:dPhiav}
  \left\langle
    \delta\Phi_{g}^{2}
  \right\rangle =
  \frac{3}{32}
  \left\langle
    h^{2}
  \right\rangle
  \Phi_{\mathrm{vac}}^{2}.
\end{equation}
In Eq.~(\ref{eq:dPhiav}), we averaged over the angles $\varphi$
and $\vartheta$ and accounted for the fact that neutrinos are ultrarelativistic.
Equation~(\ref{eq:dPhiav}) is obtained in Appendix~\ref{sec:AVERAGE};
see Eq.~(\ref{eq:dPhigav}) there.%

First, we should supply Eq.~(\ref{eq:rhoIeq}) with the initial condition.
Taking into account that $U_{0}(0)=1$ in Eq.~(\ref{eq:U0}), we
get that $\rho_{\mathrm{I}}(0)=\rho(0)$. Then, we take that $\rho_{\mathrm{I}}(0)_{11}=F_{e}$
is the initial probability ($\sim$flux) of $\nu_{e}$, $\rho_{\mathrm{I}}(0)_{22}=F_{x}=1-F_{e}$
is the initial probability ($\sim$flux) of $\nu_{x}$, and $\rho_{\mathrm{I}}(0)_{12}=\rho_{\mathrm{I}}(0)_{21}=0$,
that implies that there are no correlations between the initial fluxes
of different flavors.

Now, Eq.~(\ref{eq:rhoIeq}) can be solved analytically. The components
of $\left\langle \rho_{\mathrm{I}}\right\rangle (t)$ have the form
\begin{align}\label{eq:rhosol}
  \left\langle
    \rho_{\mathrm{I}}
  \right\rangle _{11} & =
  F_{e}+\frac{1}{2}\sin^{2}(2\theta)(F_{x}-F_{e})
  \left[
    1-\exp(-\Gamma t)
  \right],
  \nonumber
  \\
  \left\langle
    \rho_{\mathrm{I}}
  \right\rangle _{22} & =
  F_{x}-\frac{1}{2}\sin^{2}(2\theta)(F_{x}-F_{e})
  \left[
    1-\exp(-\Gamma t)
  \right],
  \nonumber
  \\
  \left\langle
    \rho_{\mathrm{I}}
  \right\rangle _{12} & =
  \left\langle
    \rho_{\mathrm{I}}
  \right\rangle _{21}=
  \frac{1}{4}\sin(4\theta)(F_{x}-F_{e})
  \left[
    1-\exp(-\Gamma t)
  \right],
\end{align}
where
\begin{equation}\label{eq:Gamma}
  \Gamma=2
  \left\langle
    \delta\Phi_{g}^{2}
  \right\rangle
  \tau =
  \frac{3}{16}
  \left\langle
    h^{2}
  \right\rangle
  \Phi_{\mathrm{vac}}^{2}\tau,
\end{equation}
is the parameter describing the relaxation of the density matrix.

The expression for $\left\langle \rho\right\rangle (t)=U_{0}\left\langle \rho_{\mathrm{I}}\right\rangle (t)U_{0}^{\dagger}$
is quite cumbersome in general case. We present its diagonal elements
in the limit $\Gamma t\gg1$,
\begin{align}\label{eq:rhodiaglim}
  \left\langle
    \rho
  \right\rangle _{11} & =
  \frac{1}{2}
  \left[
    1-\cos^{2}(2\theta)(F_{x}-F_{e})
  \right],
  \nonumber
  \\
  \left\langle
    \rho
  \right\rangle _{22} & =
  \frac{1}{2}
  \left[
    1+\cos^{2}(2\theta)(F_{x}-F_{e})
  \right].
\end{align}
One can see in Eq.~(\ref{eq:rhodiaglim}) that $\left\langle \rho\right\rangle _{11}+\left\langle \rho\right\rangle _{22}=1$,
as it should be. Of course, this relation holds true for arbitrary
$t$.

Despite the two flavors approximation, adopted above, allows the analytical solution of Eq.~\eqref{eq:rhoIeq}, it cannot correspond to a realistic situation because the mixing between $\nu_\mu$ and $\nu_\tau$ is close to maximal~\cite{Est19}. That is why we should generalize our treatment of neutrino oscillations in stochastic GWs for three flavors.

We suppose that the condition $\omega L |\beta_a - \beta_b| \ll 1$, where $a,b = 1,\dots,3$ and $L$ is the neutrino propagation distance, established in Appendix~\ref{sec:AVERAGE}, is valid. Then, performing similar calculations, as above, and using the results of Ref.~\cite{LorBal94}, we obtain the following equation for the averaged density matrix $\left\langle \rho_{\mathrm{I}}\right\rangle$ for the three neutrino flavors:
\begin{equation}\label{eq:rhoIeq3F}
  \frac{\mathrm{d}}{\mathrm{d}t}
  \left\langle
    \rho_{\mathrm{I}}
  \right\rangle = -
  \frac{3}{64}
  \langle h^2 \rangle
  \tau
  [ M, [M, \langle \rho_{\mathrm{I}} \rangle ] ],
\end{equation}
where
\begin{equation}\label{eq:Mdef3F}
  M = \frac{1}{2E} U \cdot \text{diag}
  \left(
    0, \Delta m_{21}^2, \Delta m_{31}^2
  \right)
  \cdot U^\dag.
\end{equation}
Here $ \Delta m_{ab}^2 =  m_{a}^2 - m_{b}^2$ is the standard definition for the mass squared differences. Unlike Eq.~\eqref{eq:U}, in three flavors case, the mixing matrix $U$ in Eq.~\eqref{eq:Mdef3F} can be parametrized in the form~\cite{Bil18,Est19}
\begin{equation}\label{eq:U3F}
  U =
  \left(
    \begin{array}{ccc}
      1 & 0 & 0\\
      0 & c_{23} & s_{23}\\
      0 & -s_{23} & c_{23}
    \end{array}
  \right)
  \cdot
  \left(
    \begin{array}{ccc}
      c_{13} & 0 & s_{13}e^{-\mathrm{i}\delta_\mathrm{CP}}\\
      0 & 1 & 0\\
      -s_{13}e^{\mathrm{i}\delta_\mathrm{CP}} & 0 & c_{13}
    \end{array}
  \right)
  \cdot
  \left(
    \begin{array}{ccc}
      c_{12} & s_{12} & 0\\
      -s_{12} & c_{12} & 0\\
      0 & 0 & 1
    \end{array}
  \right),
\end{equation}
where $c_{ab} = \cos\theta_{ab}$, $s_{ab} = \sin\theta_{ab}$, $\theta_{ab}$ are the corresponding vacuum mixing angles, and $\delta_\mathrm{CP}$ is the CP violating phase.

Note that Eq.~\eqref{eq:Mdef3F} cannot be solved analytically. We present its numerical solutions in Sec.~\ref{sec:ASTROPHYS} when we discuss some possible astrophysical applications.

\section{Astrophysical applications\label{sec:ASTROPHYS}}

In this section, we apply the results of Sec.~\ref{sec:FLOSCGW}
for the description of neutrino flavor oscillations in GWs emitted
by random merging binaries.

To study neutrino flavor oscillations in stochastic GWs, we should estimate $\langle h^2 \rangle$ and $\tau$ for this gravitational background. For this purpose, it is convenient to use the spectral function~\cite{Chr19},
\begin{equation}\label{eq:Omega}
  \Omega_\mathrm{GW}(f) = \frac{\pi f^3}{8G\rho_c} S_h(f),
\end{equation}
where $f = \omega/2\pi$, $G = M_\mathrm{Pl}^{-2}$ is the Newton constant, $M_\mathrm{Pl} = 1.2\times 10^{19}\,\text{GeV}$ is the Planck mass, $\rho_c = 3H_0^2/8\pi G = 0.53\times 10^{-5}\,\text{GeV}\cdot\text{cm}^{-3}$ is the critical energy density of the Universe, and $S_h(f)$ is the spectral density. The root mean square of the strain $h$ is then
\begin{equation}\label{eq:rmsh}
  \langle h^2 \rangle = \int_0^\infty \mathrm{d} f S_h(f).
\end{equation}
Instead of Eqs.~\eqref{eq:Omega} and~\eqref{eq:rmsh}, we can roughly estimate $\langle h^2 \rangle$ as
\begin{equation}\label{eq:rmsh}
  \langle h^2 \rangle \sim
  \frac{8 \rho_c \Omega_\mathrm{GW}(\tilde{f})}{\pi M_\mathrm{Pl}^{2}\tilde{f}^2},
\end{equation}
where $\tilde{f}$ is the typical frequency of stochastic GWs.

Despite merging BHs with several solar masses or neutron stars are more abundant sources of GWs, we study coalescing supermassive BHs, with masses up to $10^{10}M_\odot$. Such sources of stochastic GWs have lower characteristic frequencies $\tilde{f}$. Thus, $\langle h^2 \rangle$ is greater for such sources. In this situation, we can take $\Omega_\mathrm{GW}(\tilde{f}) = 10^{-8}$ at $\tilde{f} = 10^{-6}\,\text{Hz}$~\cite{Ros11} to get the upper limit for $\langle h^2 \rangle$. Using Eq.~\eqref{eq:rmsh}, we obtain that $\langle h^2 \rangle = 1.6\times 10^{-32}$. To evaluate the correlation time we take that $\tau \sim \tilde{f}^{-1} = 10^{6}\,\text{s}$.

%

Using the above estimates, we get the parameter, which describes the rate of the relaxation
of the probabilities in Eqs.~(\ref{eq:rhosol}) and~(\ref{eq:Gamma}), while a relativistic neutrino passes the distance $L = t$,
in the form
\begin{equation}\label{eq:kappaparam}
  \varkappa=\Gamma L=\frac{3}{16}
  \left\langle
    h^{2}
  \right\rangle
  \Phi_{\mathrm{vac}}^{2}\tau L.
\end{equation}
If $\varkappa>1$, the probabilities to detect $\nu_{e,x}$ are given
in Eq.~(\ref{eq:rhodiaglim}) since $\exp(-\varkappa)\approx0$ in
Eq.~(\ref{eq:rhosol}).


As we mentioned above, we study neutrino flavor oscillations in stochastic GWs emitted by supermassive BHs. Basing on Eq.~(\ref{eq:kappaparam}), we get that, for $\nu_{e}\to\nu_{\mu}$
oscillations channel with $\Delta m_{\odot}^{2}=7.4\times10^{-5}\,\text{eV}^{2}$~\cite{Est19},
the parameter $\varkappa$ reads
\begin{equation}\label{eq:kappaemu}
  \varkappa_{\nu_{e}\to\nu_{\mu}}=2.4\times
  \left(
    \frac{E}{10^2\,\text{keV}}
  \right)^{-2}
  \left(
    \frac{L}{1\,\text{Gpc}}
  \right).
\end{equation}
For $\nu_{e}\to\nu_{\tau}$ oscillations with $\Delta m^{2}_{31}=2.5\times10^{-3}\,\text{eV}^{2}$~\cite{Est19},
using Eq.~(\ref{eq:kappaparam}), one has
\begin{equation}\label{eq:kappaetau}
  \varkappa_{\nu_{e}\to\nu_{\tau}}=27.4\times
  \left(
    \frac{E}{1\,\text{MeV}}
  \right)^{-2}
  \left(
    \frac{L}{1\,\text{Gpc}}
  \right).
\end{equation}
One can see in Eqs.~(\ref{eq:kappaemu}) and~(\ref{eq:kappaetau})
that, if $L\gtrsim1\,\text{Gpc}$ and/or $E\lesssim10^2\,\text{keV}$ for 
$\nu_{e}\to\nu_{\mu}$ oscillations, as well as $E\lesssim\text{several MeV}$ for $\nu_{e}\to\nu_{\tau}$ oscillations, the parameter $\varkappa>1$.

Let us suppose that the ratios of fluxes at a source are $(F_{e}:F_{\mu})_{\mathrm{S}}=(1:2)$
and $(F_{e}:F_{\tau})_{\mathrm{S}}=(1:0)$. This situation is consistent with the suggestion of Ref.~\cite{Bea03}, $(F_{e}:F_{\mu}:F_{\tau})_{\mathrm{S}}=(1:2:0)$. If $\varkappa_{\nu_{e}\to\nu_{x}}>1$
in Eqs.~(\ref{eq:kappaemu}) and~(\ref{eq:kappaetau}), then, using Eq.~(\ref{eq:rhodiaglim}),
we predict the ratios of the fluxes in a terrestrial detector in the form,
\begin{align}\label{eq:fluxdet}
  \left(
    \frac{F_{e}}{F_{\mu}}
  \right)_\oplus & =
  \frac{3-\cos^{2}(2\theta_{\odot})}{3+\cos^{2}(2\theta_{\odot})}=0.9,
  \nonumber
  \\
  \left(
    \frac{F_{e}}{F_{\tau}}
  \right)_\oplus & =
  \frac{1+\cos^{2}(2\theta_{13})}{1-\cos^2(2\theta_{13})}=21.8,
\end{align}
where we take the mixing angles from Ref.~\cite{Est19}. Note that
Eq.~(\ref{eq:fluxdet}) is based on two flavors approximation $\nu_{e}\to\nu_{\mu}$
and $\nu_{e}\to\nu_{\tau}$, i.e. we do not take into account $\nu_{\mu}\leftrightarrow\nu_{\tau}$
transitions. Nevertheless, we can see that $(F_{e}:F_{\mu})_\oplus\neq(1:1)$
and $(F_{e}:F_{\tau})_\oplus\neq(1:1)$.

Now we turn to the discussion of the general three flavors situation. It is convenient to rewrite Eqs.~\eqref{eq:rhoIeq3F} and~\eqref{eq:Mdef3F} in the dimensionless form,
\begin{equation}\label{eq:rhoIeq3Fmod}
  \frac{\mathrm{d}}{\mathrm{d}t'}
  \left\langle
    \rho_{\mathrm{I}}
  \right\rangle = -
  K [ \tilde{M}, [\tilde{M}, \langle \rho_{\mathrm{I}} \rangle ] ],
\end{equation}
where $\tilde{M} = U\cdot\text{diag}(0,\Delta m_{21}^{2}/\Delta m_{31}^{2},1)\cdot U^{\dagger}$, $t' = t/L$ and
\begin{equation}\label{eq:K}
  K = \frac{3}{64}
  \left\langle
    h^{2}
  \right\rangle
  \left(
    \frac{\Delta m_{31}^{2}}{2E}
  \right)^{2}
  \tau L =
  1.8\times10^{8}
  \left(
    \frac{\Delta m_{31}^{2}}{1\,\text{eV}^{2}}
  \right)^{2}
  \left(
    \frac{E}{1\,\text{MeV}}
  \right)^{-2}
  \left(
    \frac{L}{1\,\text{Gpc}}
  \right).
\end{equation}
The dimensionless time $0<t'<1$ since $t<L$.

The estimates of the correlation time $\tau$ and $\langle h^2 \rangle$ are made in the same manner as in the two flavors case above. Should one have the numerical solution of Eq.~\eqref{eq:rhoIeq3Fmod}, the observed fluxes of flavor neutrinos can be obtained as the diagonal elements of $\left\langle \rho\right\rangle (t) =U_{0}\left\langle \rho_{\mathrm{I}}\right\rangle (t)U_{0}^{\dagger}$, where $U_{0}=\exp \left( -\mathrm{i}H_{0}t \right)$ [see Eq.~\eqref{eq:U0}] and $H_0$ is the $3\times 3$ effective Hamiltonian for flavor oscillations in vacuum.

\begin{figure}
  \centering
  \subfigure[]
  {\label{2a}
  \includegraphics[scale=.35]{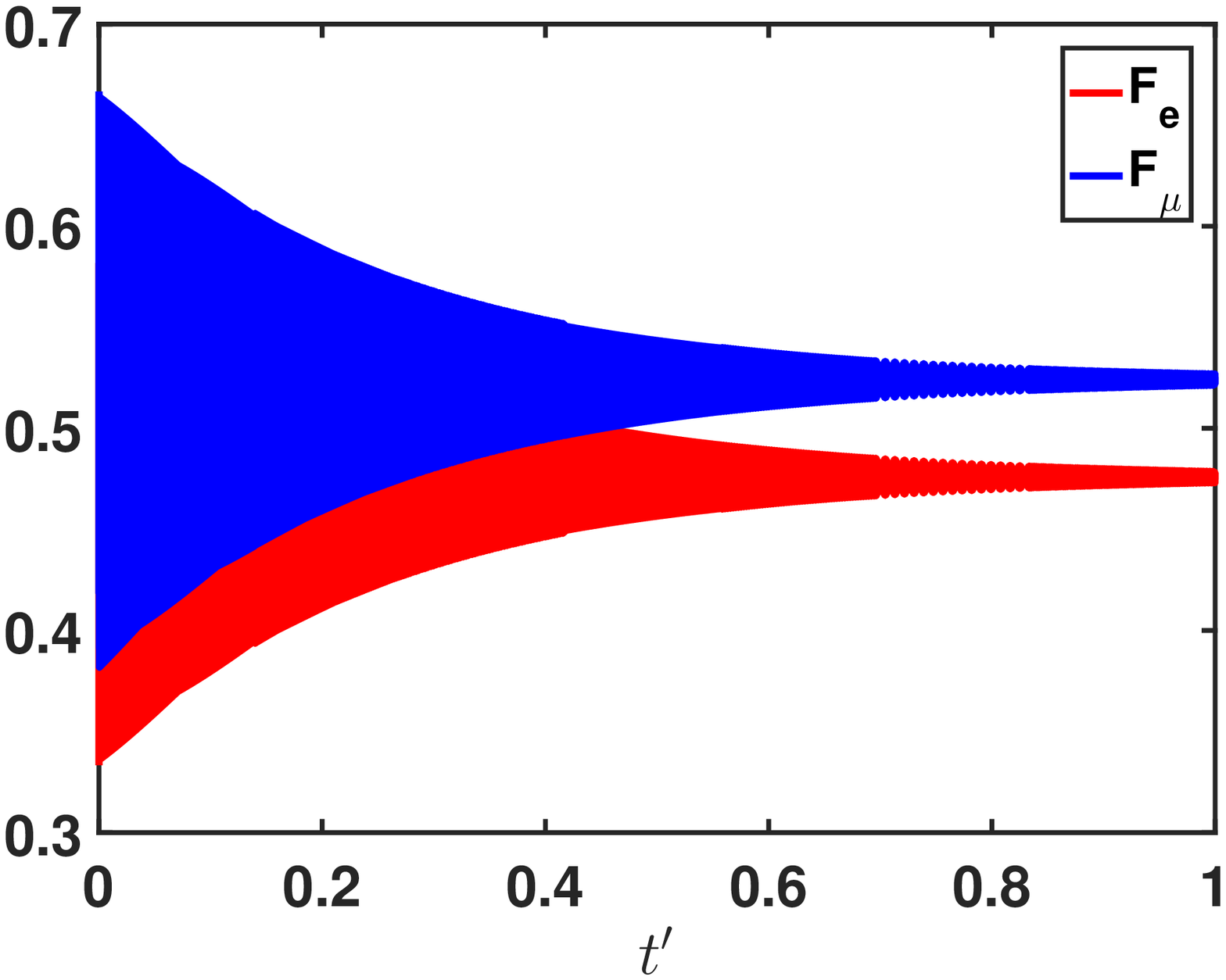}}
  \hskip-.6cm
  \subfigure[]
  {\label{2b}
  \includegraphics[scale=.35]{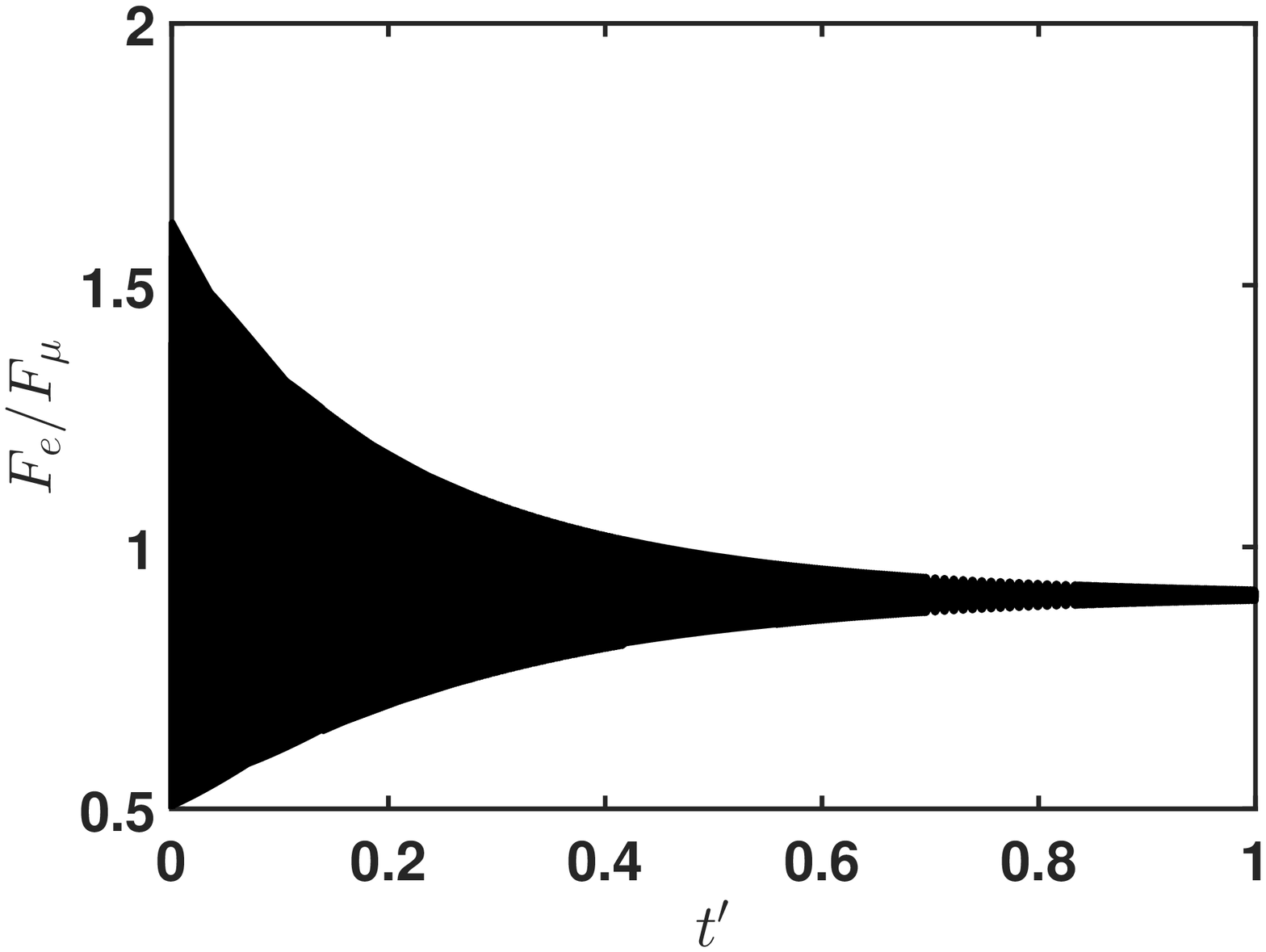}}
  \protect
  \caption{The  numerical solution
  of Eqs.~\eqref{eq:rhoIeq3Fmod} and~\eqref{eq:K}, rewritten in the two flavors approximation, 
  for $\nu_e \leftrightarrow \nu_\mu$ oscillations
  in stochastic GWs.   
  The propagation distance $L = 1\,\text{Gpc}$,
  $\Delta m_{21}^2 = \Delta m_{\odot}^2 = 7.39\times10^{-5}\,\text{eV}^2$,
  $\theta_{12} = \theta_{\odot} = 0.59$, and $E = 0.5\,\text{MeV}$.
  (a) The fluxes of electron neutrinos $F_e = \langle \rho_{11} \rangle$ (red line)
  and muon neutrinos $F_\mu = \langle \rho_{22} \rangle$ (blue line)
  versus the dimensionless time $t' = t/L$.
  (b) The ratio of fluxes $F_e/F_\mu$ as a function of $t'$.
  \label{fig:fluxes2F}}
\end{figure}

Before we proceed with the studies of the three flavors evolution, it is necessary to repeat the analytical result in Eq.~\eqref{eq:fluxdet} using the direct numerical solution of Eqs.~\eqref{eq:rhoIeq3Fmod} and~\eqref{eq:K}, rewritten for two flavors. The reduction to the two flavors case is straightforward. This numerical solution is represented in Fig.~\ref{fig:fluxes2F}. In Fig.~\ref{2a}, we show the fluxes of $\nu_e$ and $\nu_\mu$ versus $t'$. This solution is based on the fluxes at a source (the initial condition) in the form, $F_e(0) = 1/3$ and $F_\mu(0) = 2/3$. The characteristics of neutrinos and GW are taken as in Eq.~\eqref{eq:kappaemu}. One can see in Fig.~\ref{2b} that the asymptotic value of the fluxes ratio is $(F_e/F_\mu)_\oplus = 0.9$, which is in the agreement with Eq.~\eqref{eq:fluxdet}. Performing analogous simulations, we show that $(F_e/F_\tau)_\oplus = 21.8$ in Eq.~\eqref{eq:fluxdet} can be also reproduced.

\begin{figure}
  \centering
  \subfigure[]
  {\label{3a}
  \includegraphics[scale=.35]{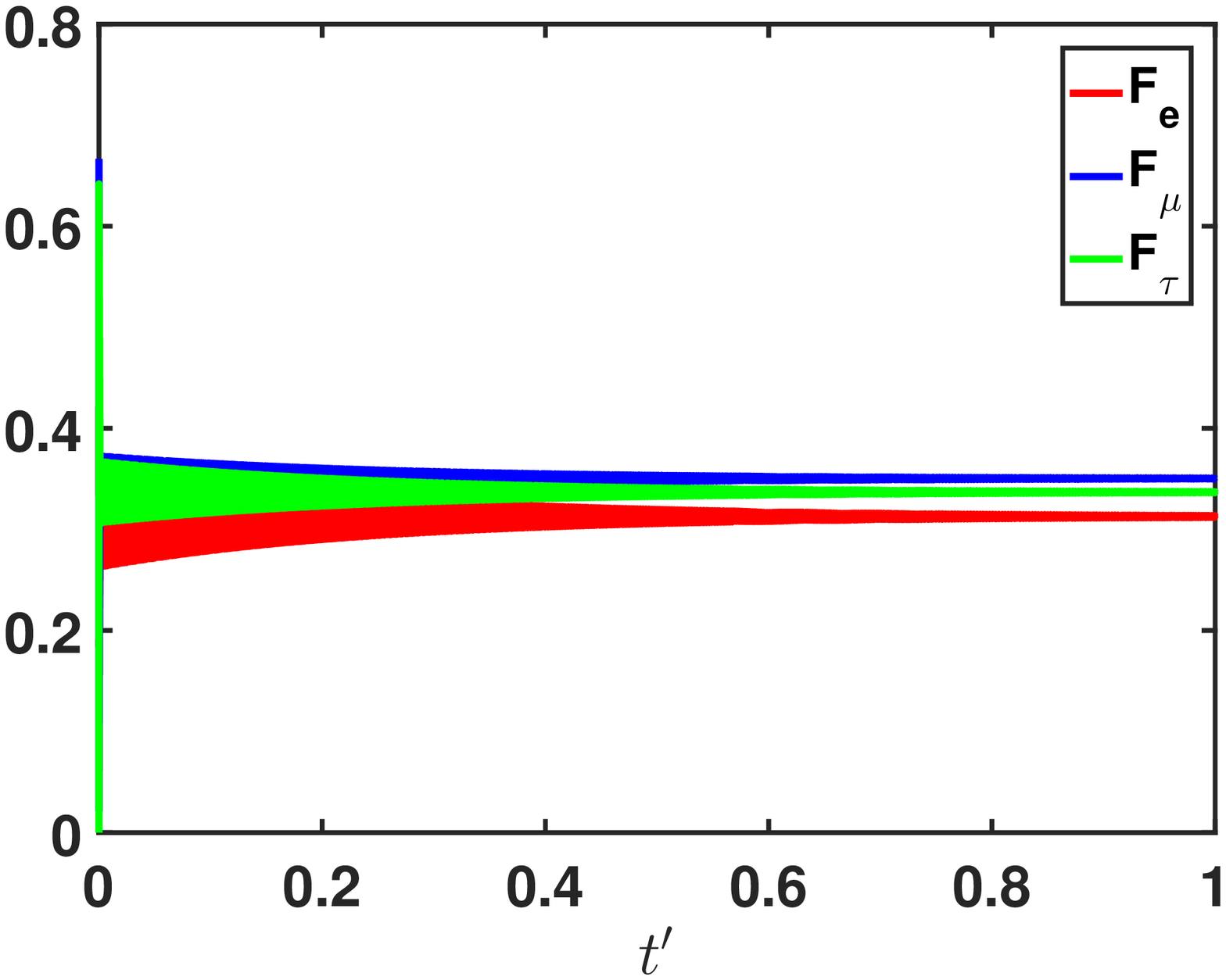}}
  \hskip-.6cm
  \subfigure[]
  {\label{3b}
  \includegraphics[scale=.35]{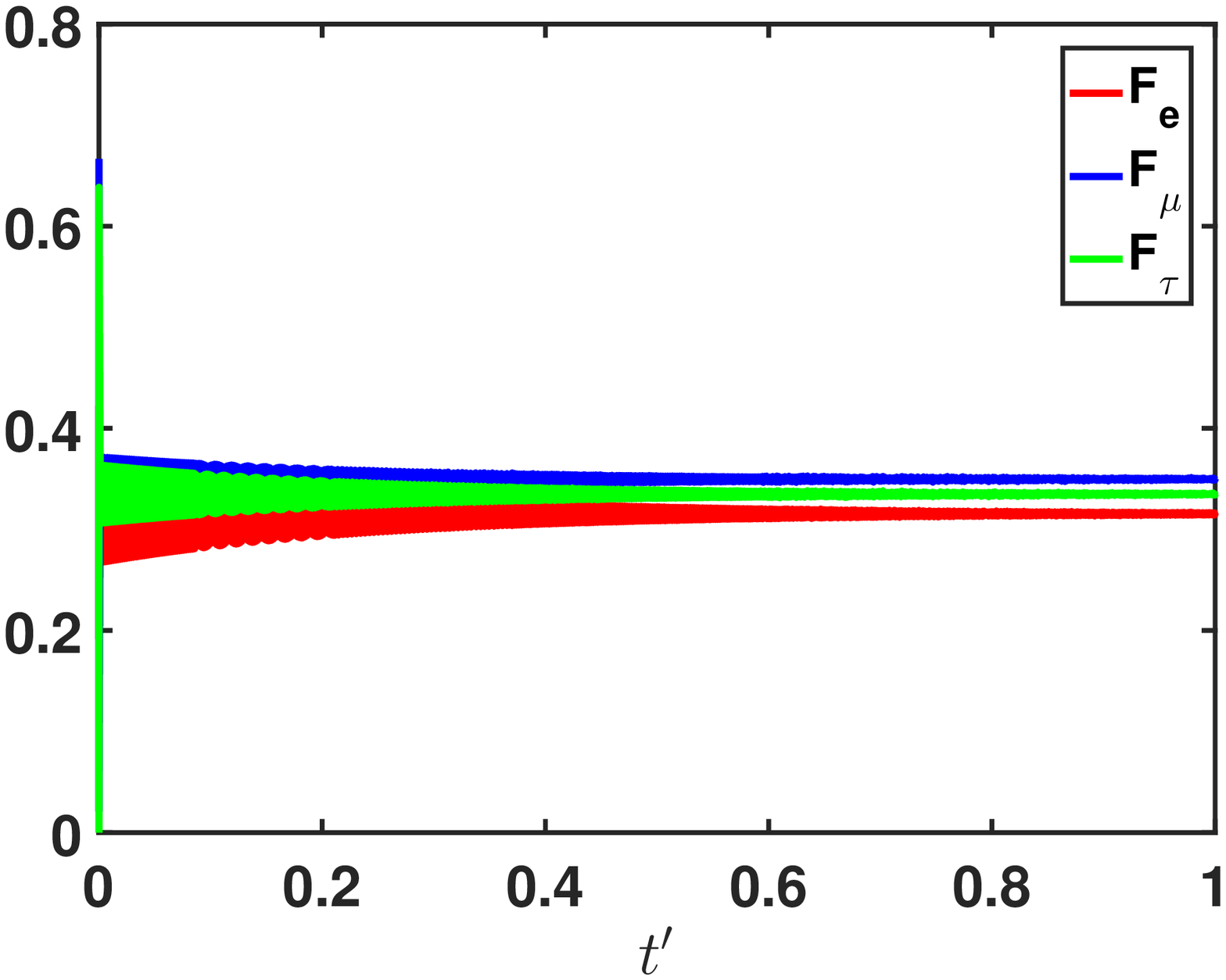}}
  \protect
  \caption{The  numerical solution
  of Eqs.~\eqref{eq:rhoIeq3Fmod} and~\eqref{eq:K} 
  for three flavor neutrinos oscillations
  in stochastic GWs. The flux of electron neutrinos
  $F_e = \langle \rho_{11} \rangle$
  versus the dimensionless time $t'$   
  is shown by the red line, the flux of muon neutrinos $F_\mu = \langle \rho_{22} \rangle$
  is represented by the blue line, and the flux of tau neutrinos $F_\tau = \langle \rho_{33} \rangle$
  is depicted by the green line.
  The neutrino energy $E=0.5\,\text{MeV}$ and the propagation distance $L=1\,\text{Gpc}$.
  (a) Normal ordering with
  $\Delta m_{21}^2 = 7.39\times10^{-5}\,\text{eV}^2$,
  $\Delta m_{31}^2 = 2.53\times10^{-3}\,\text{eV}^2$,
  $\theta_{12} = 0.59$, $\theta_{23} = 0.87$, $\theta_{13} = 0.15$,
  and $\delta_\mathrm{CP} = 4.83$;
  (b) Inverted ordering with
  $\Delta m_{21}^2 = 7.39\times10^{-5}\,\text{eV}^2$,
  $\Delta m_{31}^2 = -2.51\times10^{-3}\,\text{eV}^2$,
  $\theta_{12} = 0.59$, $\theta_{23} = 0.87$, $\theta_{13} = 0.15$,
  and $\delta_\mathrm{CP} = 4.87$.
  \label{fig:fluxes3F}}
\end{figure}

After reproducing the evolution of the two neutrino flavors with the numerical code, we can study the most general situation of three flavor neutrinos. The fluxes of $\nu_{e,\mu,\tau}$, based on the numerical solution of Eqs.~\eqref{eq:rhoIeq3Fmod} and~\eqref{eq:K}, are shown in Fig.~\ref{fig:fluxes3F}. To build Fig.~\ref{fig:fluxes3F} we utilize the values of $\Delta m_{ab}^2$, $\theta_{ab}$, and $\delta_\mathrm{CP}$ from Ref.~\cite{Est19}. We consider both normal and inverted mass hierarchies in our simulations.

In Fig.~\ref{fig:fluxes3F}, we take $L = 1\,\text{Gpc}$ and $E = 0.5\,\text{MeV}$ for the fluxes to reach their asymptotic values. This neutrino energy is between the values used in Eqs.~\eqref{eq:kappaemu} and~\eqref{eq:kappaetau}. The propagation length, taken in Fig.~\ref{fig:fluxes3F}, is comparable with the size of the visible Universe~\cite{GorRub11}.
Figure~\ref{fig:fluxes3F} is based on the initial condition (at a source) $(F_{e}:F_{\mu}:F_{\tau})_{\mathrm{S}}=(1:2:0)$. We can see in Fig.~\ref{3a} that, for the normal ordering, the asymptotic fluxes (at the Earth) are $F_{e\oplus} = 0.3127$, $F_{\mu\oplus} = 0.3504$, and $F_{\tau\oplus} = 0.3369$. For the inverted ordering, one has $F_{e\oplus} = 0.3154$, $F_{\mu\oplus} = 0.3497$, and $F_{\tau\oplus} = 0.3349$ in Fig.~\ref{3b}. It means that, at the Earth, the predicted fluxes are close to the case $(F_{e}:F_{\mu}:F_{\tau})_\oplus=(1:1:1)$. However, there is a small deviation from this prediction of Ref.~\cite{Bea03} for both normal and inverted mass orderings. Moreover, one can see that there is a small dependence of our results on the hierarchy of the neutrino masses.

Eventually, we can see that the deviation of the ratios of the cosmic neutrino fluxes, caused by the interaction with stochastic GWs, from the values $(F_{e}:F_{\mu}:F_{\tau})_\oplus=(1:1:1)$, predicted in Ref.~\cite{Bea03}, exists in the three flavors case. However, the magnitude of such a deviation is smaller than that in the two flavors approximation, studied above. The difference between the two and three flavors cases is explained by accounting for $\theta_{23}$, which is close to $\pi/4$. Hence, in the three flavors situation, $\nu_\mu \leftrightarrow \nu_\tau$  oscillations are more intense and the total neutrino flux becomes more uniform.

It is important that we study the situation of the fixed distance between a neutrino source and a neutrino detector. The only random influence on neutrino oscillations is caused by the interaction with stochastic GWs. Some other random factors, which can influence neutrino flavor oscillations, are considered in Sec.~\ref{sec:COMP}.

The recent measurement of the flavor content of cosmic neutrinos in Ref.~\cite{Aar15} excludes the following cases: $(F_{e}:F_{\mu}:F_{\tau})_\oplus=(1:0:0)$
and $(F_{e}:F_{\mu}:F_{\tau})_\oplus=(0:1:0)$. 
Our prediction of the neutrino fluxes at a source in Fig.~\ref{fig:fluxes3F} is in the region not excluded in Ref.~\cite{Aar15}.
Of course,
neutrino energies in Ref.~\cite{Aar15}, $E>35\,\text{TeV}$, are
much higher than these considered in our work, $E=(10^{2}\,\text{keV}\div\text{several MeV})$;
cf. Eqs.~(\ref{eq:kappaemu}) and~(\ref{eq:kappaetau}), as well as Figs.~\ref{fig:fluxes2F} and~\ref{fig:fluxes3F}.  Nevertheless
there are prospects to detect cosmic neutrinos even with lower energies
(see, e.g., Ref.~\cite{Bet19}). Perhaps, the proposal for the observation of the diffused flux of cosmic neutrinos in Ref.~\cite{Abe18b} would make it possible to study the influence of GWs on neutrino oscillations.

At the end of this section, we discuss the approximation made to obtain
Eqs.~(\ref{eq:rhosol}) and~(\ref{eq:Gamma}). In Appendix~\ref{sec:AVERAGE},
we find that the expressions for the neutrino fluxes in Eq.~(\ref{eq:rhodiaglim})
are valid in the limit when $\lambda=\omega t|\beta_{1}-\beta_{2}|\ll1$.
If a neutrino is a relativistic particle, it means that $t=L$ and
\begin{equation}\label{eq:Lcrit}
  L\ll L_{\mathrm{crit}},
  \quad
  L_{\mathrm{crit}}=\frac{E}{2\omega\Phi_{\mathrm{vac}}}=
  \frac{2E^{2}}{\omega\Delta m^{2}}.
\end{equation}
Let us consider $\nu_{e}\to\nu_{\mu}$ oscillations channel with $\Delta m_{\odot}^{2}=7.4\times10^{-5}\,\text{eV}^{2}$.
Taking the neutrino energy as in Eq.~(\ref{eq:kappaemu}), $E=10^2\,\text{keV}$,
and $\omega\sim10^{-6}\,\text{s}^{-1}$~\cite{Ros11}, we get that
$L_{\mathrm{crit}}=6.5\times10^{2}\,\text{Gpc}$. One can see in Eq.~(\ref{eq:kappaemu})
that, if $L\sim 1\,\text{Gpc}$ (the size of the Universe), then this
$L$ is much less than $L_{\mathrm{crit}}$. Therefore, the constraint
in Eq.~(\ref{eq:Lcrit}) is satisfied with a large margin. Analogously
one can check that Eq.~(\ref{eq:Lcrit}) is fulfilled for $\nu_{e}\to\nu_{\tau}$
oscillations. Hence, the expression for the fluxes $\nu_{e}$ and
$\nu_{x}$, proportional to $\left\langle \rho\right\rangle _{11}$
and $\left\langle \rho\right\rangle _{22}$ in Eq.~(\ref{eq:rhodiaglim}),
are valid.

\section{Other random factors contributing neutrino flavor oscillations\label{sec:COMP}}

In this section, we study other possible random factors which, along with stochastic GWs, can contribute flavor oscillations of cosmic neutrinos.

First, we mention that neutrino interaction with randomly distributed matter can affect flavor oscillations~\cite{LorBal94,BurMic97}. On the large distances $L \sim \text{Gpc}$, studied in Sec.~\ref{sec:ASTROPHYS}, the contribution of this factor to $H_f$ in Eq.~\eqref{eq:Hf} does not exceed the following quantity:
\begin{equation}
  V_m \sim G_\mathrm{F} n_\mathrm{B} =
  G_\mathrm{F} \Omega_\mathrm{B} \frac{\rho_c}{m_p} \sim 10^{-44}\,\text{eV},
\end{equation}
where $G_\mathrm{F} = 1.17\,\text{GeV}^{-2}$ is the Fermi constant, $\Omega_\mathrm{B}=0.042$ is the barions contribution to the total energy of the Universe, and $m_p$ is the proton mass. The contribution of GW to $H_f$ is $V_\mathrm{GW} \sim h \Phi_\mathrm{vac} \sim 10^{-25}\,\text{eV}$, where we take $h = 10^{-16}$, $E = 10^2\,\text{keV}$, and consider $\nu_e\to\nu_\mu$ channel. We can see that $V_m \ll V_\mathrm{GW}$, i.e., the random matter contribution is negligible for flavor oscillations of such neutrinos.

A terrestrial detector can record a neutrino flux emitted by randomly distributed sources. We should analyze this factor in studying of flavor oscillations of cosmic neutrinos and compare it with the contribution of stochastic GWs. For this purpose, we should study neutrino flavor oscillations in vacuum and average the probabilities over the distance of the neutrino beam propagation. We analyze this case in the two flavors approximation since it allows the analytical solution.

Using Eq.~\eqref{eq:Hf} and omitting $H_1$ there, we get that probabilities to observe $\nu_{e,x}$ in a detector at the distance $L$ from a source are
\begin{equation}\label{eq:PexL}
  P_{e,x}(L) =
  \left[
    \cos^2(\Phi_\mathrm{vac}L) + \cos^2 2\theta\sin^2(\Phi_\mathrm{vac}L)
  \right]
  P_{e,x}(0)
  +\sin^2 2\theta \sin^2 (\Phi_\mathrm{vac}L) P_{x,e}(0),
\end{equation}
where $P_{e,x}(0)$ are the emission probabilities. To obtain Eq.~\eqref{eq:PexL} we assume that $\langle \nu_e(0)|\nu_x(0) \rangle =0$.

Now, we average Eq.~\eqref{eq:PexL} over $L$ by setting $\cos^2(\Phi_\mathrm{vac}L) = \sin^2(\Phi_\mathrm{vac}L) = 1/2$. Finally, we get the corresponding probabilities
\begin{equation}\label{eq:Pexav}
  \langle P_{e,x} \rangle (L) =
  \frac{1}{2}
  \left\{
    1 \mp \cos^2 2\theta [P_{x}(0) - P_{e}(0)]
  \right\},
\end{equation}
which formally coincide with these in Eq.~\eqref{eq:rhodiaglim}. This fact is not surprising. It is a consequence of the ergodic theorem~\cite{Moo15}.

Nevertheless, Eqs.~\eqref{eq:rhodiaglim} and~\eqref{eq:Pexav} correspond to completely different physical situations. In Sec.~\ref{sec:FLOSCGW}, we study the neutrino propagation between a detector and a source which are at the fixed points. The random influence on neutrino oscillations by stochastic GWs is between these points. To derive Eq.~\eqref{eq:Pexav}, we consider flavor oscillations in vacuum of neutrinos emitted by randomly distributed sources. No other stochastic influence on neutrino system is assumed now.

To differentiate between these cases, we can consider a neutrino source with adiabatically changing luminosities of different neutrino flavors. In this situation, the neutrino fluxes at the source in Eq.~\eqref{eq:rhodiaglim} are slowly varying functions of time $F_{e,x}(t)$. On the contrary, one cannot expect that $P_{e,x}(0)$ change simultaneously in all sources, especially when these sources are causally disconnected.

\section{Conclusion\label{sec:CONCL}}

In this work, we have studied neutrino flavor oscillations under the
influence of a plain GW with the circular polarization for the first time. In Sec.~\ref{sec:FLOSCGW},
we have analyzed the evolution of the mass eigenstates in the quasiclassical
approximation. Using the expression for the action for a massive particle,
interacting with GW, obtained in Ref.~\cite{Pop06}, we have derived
the contribution of GW to the effective Hamiltonian for the neutrino
mass eigenstates. We have revealed that, in case of the neutrino propagation
along GW, GW does not influence neutrino flavor oscillations.

Then, we have assumed that we deal with stochastic GWs emitted by
randomly distributed sources. In this situation, we have derived the
equation for the density matrix of neutrinos using the approach in Ref.~\cite{LorBal94}. This equation has been solved analytically
in the two flavors system. The asymptotic expressions for the diagonal
elements of the density matrix, which the fluxes of flavor neutrinos
are proportional to, have been presented in Eq.~(\ref{eq:rhodiaglim}). The equation for the density matrix evolution for the three flavor neutrinos has been also derived in Sec.~\ref{sec:FLOSCGW}; cf. Eqs.~\eqref{eq:rhoIeq3F}-\eqref{eq:U3F}. However, this equation can be solved only numerically.

Then, in Sec.~\ref{sec:ASTROPHYS}, we have considered an astrophysical
application of the obtained result. We have supposed that GWs are
emitted by merging binary supermassive BHs.
Using the two flavors approximation, we have obtained
that the probabilities to detect relatively low energy neutrinos,
with $E=(10^2\,\text{keV}\div \text{several MeV})$, can reach the asymptotic values
in Eq.~(\ref{eq:rhodiaglim}) if the propagation distance is comparable
with the size of the Universe $L\sim\text{Gpc}$~\cite{GorRub11}. Recently, we
revealed in Ref.~\cite{Dvo19} that spin oscillations can be significantly
affected by GW only if the neutrino energy is low.

Note that the predicted fluxes of different flavors are not equal at a detector, contrary to the finding of Ref.~\cite{Bea03}. In the two flavors approximation, they depend on the fluxes at a source and the vacuum mixing angle. The fluxes of flavor
neutrinos at the detector in Eq.~(\ref{eq:fluxdet}) are in a region
not excluded by the observations in Ref.~\cite{Aar15}.

In Sec.~\ref{sec:ASTROPHYS}, we have also studied neutrino flavor oscillations in stochastic GWs, emitted by merging BHs, in the most general case of the three flavors. The deviation of the fluxes of flavor neutrinos at a detector from the prediction of Ref.~\cite{Bea03}, $(F_{e}:F_{\mu}:F_{\tau})_\oplus=(1:1:1)$, owing to the neutrino interaction with stochastic GWs, has been confirmed in the three neutrinos situation by means of the numerical simulations; see Fig.~\ref{fig:fluxes3F}. This deviation turns out to be smaller than that in the two flavors approximation.


The prediction of the fluxes at a detector in Sec.~\ref{sec:ASTROPHYS} is based on the assumption that the distance between a source and a detector is fixed. The stochastic influence of external fields (GWs in our case) is applied between the emission and detection points. We have found in Sec.~\ref{sec:COMP} that the same asymptotic fluxes are obtained if one averages vacuum probabilities over the neutrino propagation distances. It is the consequence of the ergodic theorem~\cite{Moo15}. In Sec.~\ref{sec:COMP}, we have have also pointed out how to pick out these completely different physical situations.

Finally, we mention that the results obtained in the present work have some uncertainty since no stochastic GW background from supermassive BHs has been observed yet. It is mainly related to the determination of the correlation time $\tau$. The fact that GWs, emitted by merging BHs with several solar masses, have been observed~\cite{Abb18}, imposes strong constraints on the parameters of such GWs. Nevertheless there are efforts to detect stochastic GWs with $10^{-9}\,\text{Hz} < f < 10^{-6}\,\text{Hz}$ (see, e.g., Ref.~\cite{Bur19}).



\begin{acknowledgments}
I am thankful to G.~Sigl and S.~V.~Troitsky for useful comments.
This work is performed within the government assignment of IZMIRAN. I am also thankful to RFBR (Grant No. 18-02-00149a) and DAAD for a partial support.
\end{acknowledgments}

\appendix

\section{Averaging of the oscillations phase induced by a gravitational wave\label{sec:AVERAGE}}

In this appendix, we obtain the expression for $\left\langle \delta\Phi_{g}^{2}\right\rangle $.
For this purpose, we should average this quantity over the angles $\vartheta$
and $\varphi$,
\begin{equation}
  \left\langle
  \delta\Phi_{g}^{2}
  \right\rangle =
  \frac{
  \left\langle
    h^{2}
  \right\rangle p^{4}}{8}\int_{0}^{\pi}
  \frac{\mathrm{d}\vartheta}{\pi}\sin^{4}\vartheta\int_{0}^{2\pi}
  \frac{\mathrm{d}\varphi}{2\pi}
  \left(
    \frac{\cos\alpha_{1}}{E_{1}}-\frac{\cos\alpha_{2}}{E_{2}}
  \right)^{2},
\end{equation}
where $\alpha_{1,2}=2\varphi-\phi_{1,2}=2\varphi-\omega t(1-\beta_{1,2}\cos\vartheta)$.
We take $t=t_{1}=t_{2}$ in the definition of $\alpha_{1,2}$ since
$\left\langle h(t_{1})h(t_{2})\right\rangle =2\tau\delta(t_{1}-t_{2})$.

The integral over $\varphi$ reads
\begin{equation}
  \int_{0}^{2\pi}\frac{\mathrm{d}\varphi}{2\pi}
  \left(
    \frac{\cos\alpha_{1}}{E_{1}}-\frac{\cos\alpha_{2}}{E_{2}}
  \right)^{2}=
  \frac{1}{2}
  \left\{
    \frac{1}{E_{1}^{2}}+\frac{1}{E_{2}^{2}}
    -2\frac{\cos[\omega t(\beta_{1}-\beta_{2})\cos\vartheta]}{E_{1}E_{2}}
  \right\} .
\end{equation}
Therefore we have
\begin{equation}\label{eq:deltaPhi}
  \left\langle
    \delta\Phi_{g}^{2}
  \right\rangle =
  \frac{3
  \left\langle
    h^{2}
  \right\rangle
  p^{4}}{128}
  \left(
    \frac{1}{E_{1}^{2}}+\frac{1}{E_{2}^{2}}-2\frac{I(\lambda)}{E_{1}E_{2}}
  \right),
  \quad
  I(\lambda) =
  \frac{8}{3\pi}\int_{0}^{\pi}\mathrm{d}\vartheta\sin^{4}\vartheta\cos(\lambda\cos\vartheta),
\end{equation}
where $\lambda=\omega t(\beta_{1}-\beta_{2})$.

Using the fact that
\begin{equation}
  \frac{1}{\pi}\int_{0}^{\pi}\mathrm{d}\vartheta\cos(\lambda\cos\vartheta)=J_{0}(\lambda),
\end{equation}
where $J_{0}(\lambda)$ is the Bessel function, we get that
\begin{equation}
  I(\lambda)=\frac{8}{3}
  \left[
    J_{0}(\lambda)+2J_{0}^{\prime\prime}(\lambda)+J_{0}^{(\mathrm{IV})}(\lambda)
  \right]=
  \frac{8}{\lambda^{3}}
  \left[
    2J_{1}(\lambda)-\lambda J_{0}(\lambda)
  \right].
\end{equation}
Here, $J_{1}(\lambda)$ is the Bessel function.

If $|\lambda|\ll1$, $I(\lambda)\to1$. Thus, $\left\langle \delta\Phi_{g}^{2}\right\rangle $
in Eq.~(\ref{eq:deltaPhi}) takes the form
\begin{equation}\label{eq:dPhigav}
  \left\langle
    \delta\Phi_{g}^{2}
  \right\rangle \approx
  \frac{3
  \left\langle
    h^{2}
  \right\rangle
  p^{4}}{128}
  \left(
    \frac{1}{E_{1}}-\frac{1}{E_{2}}
  \right)^{2}=
  \frac{3}{32}
  \left\langle
    h^{2}
  \right\rangle
  \Phi_{\mathrm{vac}}^{2}.
\end{equation}
Equation~(\ref{eq:dPhigav}) is used in the master Eq.~(\ref{eq:rhoIeq})
for the density matrix.


\begin{thebibliography}{50}

\bibitem{Ago18}
  M.~Agostini \textit{et al.} (Borexino Collaboration), 
  Comprehensive measurement of $pp$-chain solar neutrinos,
  Nature (London) \textbf{562}, 505\textendash 510 (2018).

\bibitem{Abe18}
  K.~Abe \textit{et al.} (Super-Kamiokande Collaboration), 
  Atmospheric neutrino oscillation analysis with external constraints
  in Super-Kamiokande I-IV,
  Phys. Rev. D \textbf{97}, 072001 (2018)
  [arXiv:1710.09126].

\bibitem{Ace18}
  M.~A.~Acero \textit{et al.} (NOvA Collaboration), 
  New constraints on oscillation parameters from $\nu_{e}$ appearance
  and $\nu_{\mu}$ disappearance in the NOvA experiment,
  Phys. Rev. D \textbf{98}, 032012 (2018)
  [arXiv:1706.04592].

\bibitem{Bil18}
  S.~Bilenky,
  \textit{Introduction to the Physics of Massive and Mixed Neutrinos}, 2nd ed.
  (Springer, Cham, 2018).

\bibitem{Giu16}
  C.~Giunti, K.~A.~Kouzakov, Y.-F.~Li, A.~V.~Lokhov, A.~I.~Studenikin, and S.~Zhou,
  Electromagnetic neutrinos in laboratory experiments and astrophysics,
  Ann. Phys. (Amsterdam) \textbf{528}, 198\textendash 215 (2016)
  [arXiv:1506.05387].

\bibitem{MalSmi16}
  M.~Maltoni and A.~Yu.~Smirnov,
  Solar neutrinos and neutrino physics,
  Eur. Phys. J. A \textbf{52}, 87 (2016)
  [arXiv:1507.05287].

\bibitem{AhlBur96}
  D.~V.~Ahluwalia and C.~Burgard,
  Gravitationally induced neutrino-oscillation phases,
  Gen. Relativ. Gravit. \textbf{28}, 1161\textendash 1170 (1996)
  [gr-qc/9603008].

\bibitem{For97}
  N.~Fornengo, C.~Giunti, C.~W.~Kim, and J.~Song,
  Gravitational effects on the neutrino oscillation,
  Phys. Rev. D \textbf{56}, 1895\textendash 1902 (1997)
  [hep-ph/9611231].

\bibitem{GodPas11}
  S.~I.~Godunov and G.~S.~Pastukhov,
  Neutrino oscillations in the gravitational field,
  Phys. Atom. Nucl. \textbf{74}, 302\textendash 305 (2011)
  [arXiv:0906.5556].

\bibitem{Vis15}
  L.~Visinelli,
  Neutrino flavor oscillations in a curved space-time,
  Gen. Relativ. Gravit. \textbf{47}, 62 (2015)
  [arXiv:1410.1523].

\bibitem{Abb16}
  B.~P.~Abbott \textit{et al.} (LIGO Scientific Collaboration and Virgo Collaboration), 
  Observation of gravitational waves from a binary black hole merger,
  Phys. Rev. Lett. \textbf{116}, 061102 (2016) 
  [arXiv:1602.03837].

\bibitem{Abb18}
  B.~P.~Abbott \textit{et al.} (LIGO Scientific Collaboration and Virgo Collaboration), 
  GWTC-1: A gravitational-wave transient catalog of compact binary mergers observed
  by LIGO and Virgo during the first and second observing runs,
  Phys. Rev. X \textbf{9}, 031040 (2019)
  [arXiv:1811.12907].

\bibitem{Alb19}
  A.~Albert \textit{et al.} (ANTARES, IceCube, LIGO, and Virgo Collaborations), 
  Search for multi-messenger sources of gravitational waves and high-energy neutrinos
  with advanced LIGO during its first observing run, ANTARES and IceCube,
  Astrophys. J. \textbf{870}, 134 (2019)
  [arXiv:1810.10693].

\bibitem{Dvo19}
  M.~Dvornikov,
  Neutrino spin oscillations in external fields in curved spacetime,
  Phys. Rev. D \textbf{99}, 116021 (2019)
  [arXiv:1902.11285].

\bibitem{Dvo06}
  M.~Dvornikov,
  Neutrino spin oscillations in gravitational fields,
  Int. J. Mod. Phys. D \textbf{15}, 1017\textendash 1034 (2006)
  [hep-ph/0601095].

\bibitem{Dvo13}
  M.~Dvornikov,
  Neutrino spin oscillations in matter under the influence of gravitational
  and electromagnetic fields,
  J. Cosmol. Astropart. Phys. 06 (2013) 015
  [arXiv:1306.2659].

\bibitem{ObuSilTer17}
  Y.~N.~Obukhov, A.~J.~Silenko, and O.~V.~Teryaev,
  General treatment of quantum and classical spinning particles in external fields,
  Phys. Rev. D \textbf{96}, 105005 (2017)
  [arXiv:1708.05601].

\bibitem{Pop06}
  N.~J.~Pop\l{}awski,
  A Michelson interferometer in the field of a plane gravitational wave,
  J. Math. Phys. (N.Y.) \textbf{47}, 072501 (2006)
  [gr-qc/0503066].

\bibitem{Buo07}
  A.~Buonanno,
  Course~1 -- Gravitational waves,
  in \textit{Particle Physics and Cosmology: The Fabric of Spacetime},
  edited by F.~Bernardeau, C.~Grojean, and J.~Dalibard  
  (Elsevier, Amsterdam, 2007), Vol.~86, pp.~3\textendash 52
  [arXiv:0709.4682].

\bibitem{Dvo18}
  M.~Dvornikov,
  Spin-flavor oscillations of Dirac neutrinos in a plane electromagnetic wave,
  Phys. Rev. D \textbf{98}, 075025 (2018)
  [arXiv:1806.08719].

\bibitem{LorBal94}
  F.~N.~Loreti and A.~B.~Balantekin,
  Neutrino oscillations in noisy media,
  Phys. Rev. D \textbf{50}, 4762\textendash 4770 (1994) 
  [nucl-th/9406003].

\bibitem{Est19}
  I.~Esteban, M.~C.~Gonzalez-Garcia, A.~Hernandez-Cabezudo, M.~Maltoni, and T.~Schwetz,
  Global analysis of three-flavour neutrino oscillations:
  Synergies and tensions in the determination of $\theta_{23}$, $\delta_{\mathrm{CP}}$,
  and the mass ordering,
  J. High Energy Phys. 01 (2019) 106
  [arXiv:1811.05487].

\bibitem{Chr19}
  N.~Christensen,
  Stochastic gravitational wave backgrounds,
  Rep. Prog. Phys. \textbf{82}, 016903 (2019)
  [arXiv:1811.08797].

\bibitem{Ros11}
  P.~A.~Rosado,
  Gravitational wave background from binary systems,
  Phys. Rev. D \textbf{84}, 084004 (2011)
  [arXiv:1106.5795].


\bibitem{Bea03}
  J.~F.~Beacom, N.~F.~Bell, D.~Hooper, S.~Pakvasa, and T.~J.~Weiler,
  Measuring flavor ratios of high-energy astrophysical neutrinos,
  Phys. Rev. D \textbf{68}, 093005 (2003) [hep-ph/0307025];
  Erratum, Phys. Rev. D \textbf{72}, 019901 (2005).

\bibitem{GorRub11}
  D.~S.~Gorbunov and V.~A.~Rubakov,
  \textit{Introduction to the Theory of the Early Universe: Hot Big Bang Theory}
  (World Scientific, Singapore, 2011), p.~8.

\bibitem{Aar15}
  M.~G.~Aartsen \textit{et al.} (IceCube Collaboration), 
  Flavor ratio of astrophysical neutrinos above 35~TeV in IceCube,
  Phys. Rev. Lett. \textbf{114}, 171102 (2015)
  [arXiv:1502.03376].

\bibitem{Bet19}
  M.~G.~Betti \textit{et al.} (PTOLEMY Collaboration), 
  Neutrino physics with the PTOLEMY project:
  Active neutrino properties and the light sterile case,
  J. Cosmol. Astropart. Phys. 07 (2019) 047
  [arXiv:1902.05508].

\bibitem{Abe18b}
  K.~Abe \textit{et al.} (Hyper-Kamiokande Proto-Collaboration), 
  Physics potentials with the second Hyper-Kamiokande detector in Korea,
  Prog. Theor. Exp. Phys. \textbf{2018}, 063C01 (2018)
  [arXiv:1611.06118].

\bibitem{BurMic97}
  C.~P.~Burgess and D.~Michaud, 
  Neutrino propagation in a fluctuating Sun,
  Ann. Phys. (N.Y.) \textbf{256}, 1--38 (1997)
  [hep-ph/9606295].

\bibitem{Moo15}
  C.~C.~Moore, 
  Ergodic theorem, ergodic theory, and statistical mechanics,
  Proc. Natl. Acad. Sci. U.S.A. \textbf{112}, 1907--1911 (2015).

\bibitem{Bur19}
  S.~Burke-Spolaor \textit{et al.}, 
  The astrophysics of nanohertz gravitational waves,
  Astron. Astrophys. Rev. \textbf{27}, 5 (2019)
  [arXiv:1811.08826].


%

\end{thebibliography}
\end{document}